\documentclass[onecolumn,11pt]{article}
\usepackage[top=1in, bottom=1in, left=1in, right=1in]{geometry}
\usepackage{amsmath,amsfonts,amscd,amssymb}
\usepackage{graphicx}
\usepackage{epstopdf}
\usepackage{overpic}
\usepackage{cancel}
\usepackage{rotating}
\usepackage{url}
\usepackage{caption}
\usepackage{color}
\usepackage{rotating}
\usepackage{multirow}
\usepackage{wrapfig}
\usepackage{mathtools}
\usepackage{subeqnarray}
\usepackage{setspace}
\usepackage{palatino} 
\usepackage[numbers,sort&compress]{natbib}
\usepackage{algorithm}
\usepackage{algpseudocode}

\usepackage[bottom,flushmargin,hang,multiple]{footmisc}
\usepackage{lipsum}
\newcommand\blfootnote[1]{%
  \begingroup
  \renewcommand\thefootnote{}\footnote{#1}%
  \addtocounter{footnote}{-1}%
  \endgroup
}

\newcommand{\beginsupplement}{%
        \setcounter{section}{0}
        \renewcommand{\thesection}{S\arabic{section}}%
        \setcounter{table}{0}
        \renewcommand{\thetable}{S\arabic{table}}%
        \setcounter{figure}{0}
        \renewcommand{\thefigure}{S\arabic{figure}}%
     }

\newcommand{\Cv}{\mathbf{C}}

\newcommand{\Hv}{\mathbf{H}}
\newcommand{\Iv}{\mathbf{I}}

\newcommand{\Sv}{\mathbf{S}}
\newcommand{\Xv}{\mathbf{X}}
\newcommand{\Wv}{\mathbf{W}}

\newcommand{\dv}{\mathbf{d}}
\newcommand{\rv}{\mathbf{r}}
\newcommand{\vv}{\mathbf{v}}
\newcommand{\valpha}{\boldsymbol{\alpha}}

\newcommand{\Thetav}{\boldsymbol{\Theta}}
\newcommand{\Xiv}{\boldsymbol{\Xi}}
\newcommand{\fv}{\mathbf{f}}
\newcommand{\gv}{\mathbf{g}}
\newcommand{\xv}{\mathbf{x}}

\newcommand{\alphav}{\boldsymbol{\alpha}}
\newcommand{\phiv}{\boldsymbol{\phi}}

\newcommand{\thetav}{\boldsymbol{\theta}}
\newcommand{\xiv}{\boldsymbol{\xi}}
\newcommand{\zetav}{\boldsymbol{\zeta}}

\newcommand\ddt{\frac{\mathrm{d}}{\mathrm{d}t}}
\newcommand\prox{\mathrm{prox}}

\newcommand\Rb{\mathbb{R}}
\DeclareMathOperator{\dist}{dist}

\DeclareMathOperator{\proj}{proj}

\DeclareMathOperator*{\argmin}{argmin}

\newtheorem{theorem}{Theorem}

\title{A unified sparse optimization framework to learn parsimonious physics-informed models from data}
\author{\normalsize{Kathleen Champion$^{1*}$, Peng Zheng$^1$, Aleksandr Y.~Aravkin$^1$, Steven L. Brunton$^{2,1}$, J. Nathan Kutz$^1$}\\
\footnotesize{$^1$ Department of Applied Mathematics, University of Washington, Seattle, WA 98195, United States}\\
\footnotesize{$^2$ Department of Mechanical Engineering, University of Washington, Seattle, WA 98195, United States\vspace{-.2in}}
}
\date{}

\begin{document}
\maketitle

\blfootnote{$^*$ Corresponding author: (kpchamp@uw.edu).}

\begin{abstract}
Machine learning (ML) is redefining what is possible in data-intensive fields of science and engineering. However, applying ML to problems in the physical sciences comes with a unique set of challenges: scientists want physically interpretable models that can (i) generalize to predict previously unobserved behaviors, (ii) provide effective forecasting predictions (extrapolation), and (iii) be certifiable. Autonomous systems will necessarily interact with changing and uncertain environments, motivating the need for models that can accurately extrapolate based on physical principles (e.g. Newton's universal second law for classical mechanics, $F=ma$). Standard ML approaches have shown impressive performance for predicting dynamics in an interpolatory regime, but the resulting models often lack interpretability and fail to generalize.  We introduce a unified sparse optimization framework that learns governing dynamical systems models from data, selecting relevant terms in the dynamics from a library of possible functions. The resulting models are parsimonious, have physical interpretations, and can generalize to new parameter regimes. Our framework allows the use of non-convex sparsity promoting regularization functions and can be adapted to address key challenges in scientific problems and data sets, including outliers, parametric dependencies, and physical constraints. 
We show that the approach  discovers parsimonious dynamical models on several example systems. This flexible approach can be tailored to the unique challenges associated with a wide range of applications and data sets, providing a powerful ML-based framework for learning governing models for physical systems from data.
\end{abstract}

\section{Introduction}
\label{sec:introduction}
With abundant data being generated across scientific fields, researchers are increasingly turning to machine learning (ML) methods to aid scientific inquiry.  
In addition to standard techniques in clustering and classification, ML is now being used to discover models that  characterize and predict the behavior of physical systems.  
Unlike many applications in ML, interpretation, generalization and extrapolation are the primary objectives for engineering and science, hence we must identify parsimonious models that have the fewest terms required to describe the dynamics.  This is in contrast to neural networks (NNs), which are defined by exceedingly large parametrizations  which typically lack interpretability or generalizability.  
A breakthrough approach in model discovery used symbolic regression to learn the form of governing equations from data~\cite{bongard_automated_2007,schmidt_distilling_2009}. 
{\em Sparse identification of nonlinear dynamics} (SINDy)~\cite{brunton_discovering_2016} is a related approach that uses sparse regression to find the fewest terms in a library of candidate functions required to model the dynamics. 
Because this approach is based on a sparsity-promoting linear regression, it is possible to incorporate partial knowledge of the physics, such as symmetries, constraints, and conservation laws (e.g., conservation of mass, momentum, and energy)~\cite{Loiseau2017jfm}. 
In this work, we develop a unified sparse optimization framework for dynamical system discovery that enables one to simultaneously discover models, trim corrupt training data, enforce known physics, and identify parametric dependency in the equations. 

Although studied for decades in dynamical systems~\cite{gonzalez1998identification,Milano2002jcp}, the universal approximation capabilities of NNs~\cite{hornik1989multilayer,goodfellow2016deep} have generated a resurgence of approaches to model time-series data~\cite{Mardt2017arxiv,vlachas_data-driven_2018,wehmeyer2018time,yeung_learning_2017,Takeishi2017nips,lusch2018deep,raissi2017physics2,raissi2018multistep,bar2018data,pathak2018model,champion2019data}.  
NNs can also learn coordinate transformations that simplify the dynamical system representation~\cite{Mardt2017arxiv,wehmeyer2018time,yeung_learning_2017,Takeishi2017nips,lusch2018deep,MJKW18,champion2019data} (e.g. Koopman representations~\cite{koopman_hamiltonian_1931,mezic_spectral_2005}).
However, NNs generally struggle with extrapolation, are difficult to interpret, and cannot readily enforce known or partially known physics. 

In contrast, the SINDy algorithm~\cite{brunton_discovering_2016,champion2019data} has been shown to produce interpretable and generalizable dynamical systems models from limited data.  
SINDy has been applied broadly to identify models for optical systems~\cite{Sorokina2016oe}, fluid flows~\cite{Loiseau2017jfm}, chemical reaction dynamics~\cite{Hoffmann2018arxiv}, plasma convection~\cite{Dam2017pf}, structural modeling~\cite{lai2019sparse}, and for model predictive control~\cite{Kaiser2018prsa}.  
It is also possible to extend SINDy to identify partial differential equations~\cite{Rudy2017sciadv,Schaeffer2017prsa} and dynamical systems with parametric dependencies~\cite{rudy2019data}, to trim corrupt data~\cite{Tran2016mms}, and to incorporate partially known physics and constraints~\cite{Loiseau2017jfm}.  
Because the approach is fundamentally based on a sparsity-regularized regression, there is an opportunity to unify these innovations via the {\em sparse relaxed regularized regression} (SR3) algorithm~\cite{zheng2019unified}, resulting in a unified sparse model discovery framework. 

\begin{figure}
\centering
\includegraphics[width=\linewidth]{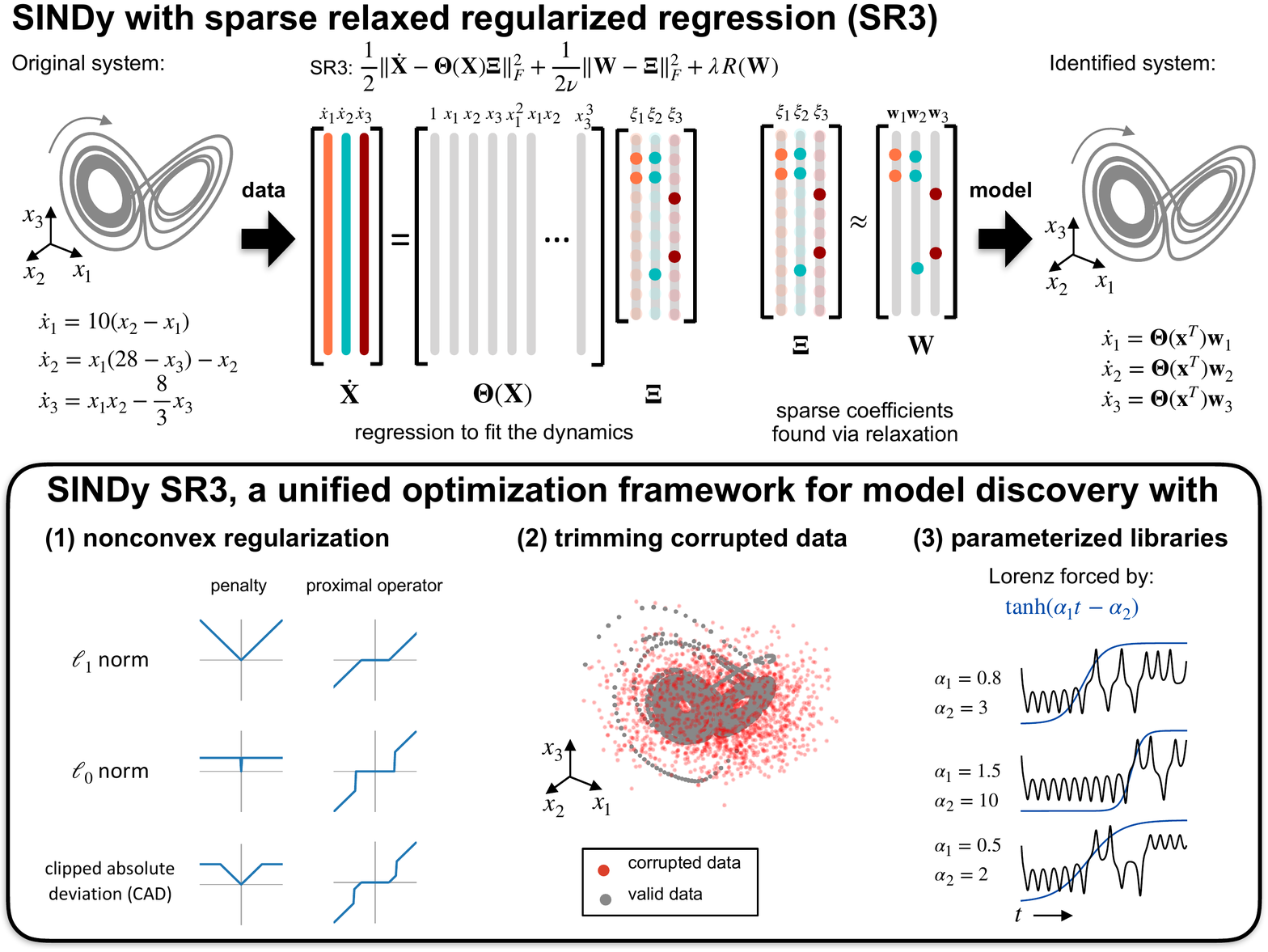}
\caption{Overview of the SINDy method for identifying nonlinear dynamical systems. SINDy sets up the system identification problem as a sparse regression problem, selecting a set of active governing terms from a library. Sparse relaxed regularized regression (SR3) provides a flexible, unified framework that can be adapted to address a number of challenges that might occur with data from physical systems, including outlier identification, parameterized library functions, and forcing.}
\label{fig:overview}
\end{figure}

\subsection{Basic problem formulation}

SINDy~\cite{brunton_discovering_2016} enables the discovery of nonlinear dynamical systems models from data. Assume we have data from a dynamical system 
\begin{equation}
  \ddt\xv(t) = \fv(\xv(t))
  \label{eq:dynamical_system_x}
\end{equation}
where $\xv(t) \!\in\! \Rb^n$ is the state of the system at time $t$. We want to find the terms in $\fv$ given the assumption that $\mathbf{f}$ has only a few active terms: it is sparse in the space of all possible functions of $\mathbf{x}(t)$. Given snapshot data $\mathbf{X} = \begin{bmatrix}\mathbf{x}_1 & \mathbf{x}_2 & \cdots & \mathbf{x}_m\end{bmatrix}^T$ and $\mathbf{\dot{X}} = \begin{bmatrix}\mathbf{\dot{x}}_1 & \mathbf{\dot{x}}_2 & \cdots & \mathbf{\dot{x}}_m\end{bmatrix}^T$,
we build a library of candidate functions $\Thetav(\Xv) = [\thetav_1(\Xv) \cdots \thetav_p(\Xv)]$. We then seek a solution of
\begin{equation*}
  \dot{\Xv} = \Thetav(\Xv)\Xiv.
\end{equation*}
where $\Xiv \!=\! (\xiv_1\ \xiv_2\ \cdots\ \xiv_n )$ are sparse coefficient (loading) vectors.
A natural optimization is given by 
\begin{equation}
\label{eq:sindy}
    \min_{\Xiv} \frac{1}{2}\| \dot{\Xv} - \Thetav(\Xv)\Xiv \|^2 + \lambda R(\Xiv)
\end{equation}
where $R(\cdot)$ is a regularizer that promotes sparsity.
When $R$ is convex, a range of well-known algorithms for~\eqref{eq:sindy} are available. The standard approach is to choose $R$ to be the sparsity-promoting $\ell_1$ norm, which is the convex relaxation of the $\ell_0$ norm. In this case,  SINDy is solved via LASSO~\cite{tibshirani2015statistical}. In practice, LASSO does not perform well at coefficient selection (see Section~\ref{sec:method_comparison} for details). In the context of dynamics discovery we would like to use non-convex $R$, specifically the $\ell_0$ norm.

The standard SINDy algorithm performs sequential thresholded least squares (STLSQ): given a parameter $\eta$ that specifies the minimum magnitude for a coefficient in $\Xiv$, perform a least squares fit and then zero out all coefficients with magnitude below the threshold. This process of fitting and thresholding is performed until convergence. While this method works very well, it is customized to the least squares 
formulation and does not readily accommodate extensions including incorporation of additional constraints, robust formulations, or nonlinear parameter estimation.   A number of extensions to SINDy have been developed and required adaptations to the optimization algorithm~\cite{Tran2016mms,Rudy2017sciadv,Schaeffer2017prsa,Loiseau2017jfm,Kaiser2018prsa}.

\section{Formulation and approach}

We extend the optimization formulation~\eqref{eq:sindy} to include additional structure,  robustness to outliers, and nonlinear parameter estimation using the sparse relaxed regularized regression (SR3) approach that uses relaxation and partial minimization~\cite{zheng2019unified}. 
SR3 for~\eqref{eq:sindy} introduces the auxiliary variable $\Wv$ and relaxes the optimization to
\begin{equation}
\label{eq:explicit_form}
    \min_{\Xiv,\Wv} \frac{1}{2}\| \dot{\Xv} - \Thetav(\Xv)\Xiv \|^2 + \lambda R(\Wv) + \frac{1}{2\nu}\|\Xiv - \Wv\|^2.
\end{equation}
We can solve~\eqref{eq:explicit_form} using the alternating update rule in Algorithm~\ref{ALG:BASIC}.  This requires only least squares solves and prox operators~\cite{zheng2019unified}. The resulting solution approximates the original problem~\eqref{eq:sindy} as $\nu \downarrow 0$. When $R$ is  taken to be the $\ell_0$ penalty, the prox operator is hard thresholding, and Algorithm~\ref{ALG:BASIC} is similar, but not equivalent, to thresholded least squares, and performs 
similarly in practice. However, unlike thresholded least squares, the SR3 approach easily generalizes to new problems and features.

\begin{algorithm}[t]
   \caption{Basic SR3 Algorithm}
   \label{ALG:BASIC}
\begin{algorithmic}
  \State{Input $\epsilon$, $\Wv^0$}
  \State{Initialize $k=0$,  err$=2\epsilon$.}
   \While{err$>\epsilon$}
   \State{$k \leftarrow k+1$}
   \State{$  \Xiv^{k}  = \displaystyle\argmin_{\Xiv} \frac{1}{2}\| \dot{\Xv} - \Thetav(\Xv)\Xiv \|^2 +  \frac{1}{2\nu}\|\Xiv - \Wv^{k-1}\|^2$}
   \State{$ \Wv^k  =  \prox_{\lambda\nu R} (\Xiv^k)$}
   \State{err $=\|\Wv^k - \Wv^{k-1}\|/\nu$}
   \EndWhile
\end{algorithmic}
\end{algorithm}

\subsection{Performance of SR3 for SINDy}\label{sec:method_comparison}
SR3 for SINDy provides an optimization framework that both (i) enables the identification of truly sparse models and (ii) can be adapted to include additional features. 
We first compare SR3 to both STLSQ and the LASSO algorithm. 
While STLSQ works well for identifying sparse models that capture the behavior of a system, it is a standalone method without a true optimization cost function, meaning the algorithm must be reformulated to work with other adaptations to the SINDy problem \cite{Mangan2016ieee}. 
LASSO provides a standard optimization approach but does not successfully identify sparse models. Even with noiseless (clean) data, LASSO models for SINDy typically have many coefficients that are small in magnitude but nonzero. Obtaining a sparse set of coefficients is key for interpretability. SR3 works with nonconvex regularization functions such as the $\ell_0$ norm, enabling the identification of truly sparse models.

\begin{figure}
\centering
\includegraphics[width=\linewidth]{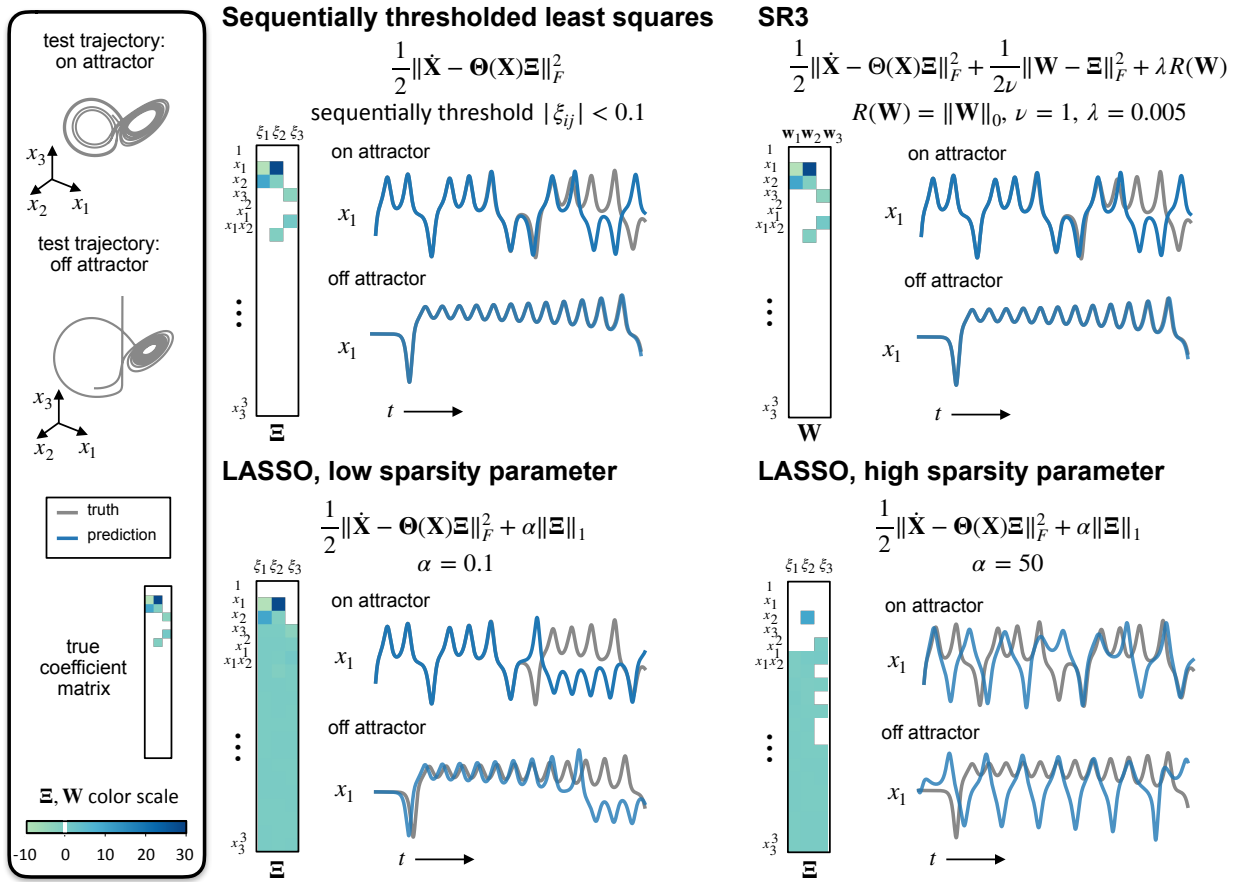}
\caption{Comparison of optimization methods for identifying the active coefficients in a SINDy model. The standard approach has been STLSQ, which is able to identify a sparse model that fits the data well. However, this approach lacks flexibility and is not easily adapted to incorporate other optimization challenges. LASSO is a standard approach for performing a sparse regression, but does not do well at performing coefficient selection: many of the terms in the coefficient matrix are small but nonzero. Increasing the regularization strength leads to a model that is still not sparse and has a poor fit of the data. SR3 relaxes the regression problem in a way that enables the use of nonconvex regularization functions such as the $\ell_0$ norm or hard thresholding. This results in a truly sparse model, and provides a flexible framework that can easily incorporate additional optimizations such as trimming outliers and fitting parameterized library functions.}
\label{fig:method_comparison}
\end{figure}

In Fig.~\ref{fig:method_comparison} we compare these algorithms using data from the canonical chaotic Lorenz system:
\begin{align}
    \dot{x}_1 &= 10 (x_2 - x_1), \nonumber \\
    \dot{x}_2 &= x_1(28 - x_3) - x_2 , \label{eq:lorenz} \\
    \dot{x}_3 &= x_1 x_2 - ({8}/{3}) x_3. \nonumber
\end{align}
We simulate the system from 20 initial conditions and fit a SINDy model with polynomials up to order 3 using the following optimization approaches: STLSQ with threshold $0.1$, SR3 with $\ell_0$ regularization, LASSO with a regularization weight of $0.1$, and LASSO with a regularization weight of $50$. For each model we analyze both the sparsity pattern of the coefficient matrix, and simulations of the resulting model on test trajectories. 
%
As shown in Fig.~\ref{fig:method_comparison}, STLSQ and SR3 yield the same correct sparsity pattern. In simulation, both track a Lorenz test trajectory for several trips around the attractor before eventually diverging. The eventual deviation is expected due to the chaotic nature of the Lorenz system, as a slight difference in coefficient values or initial conditions can lead to vastly different trajectories (although the trajectories continue to fill in the Lorenz attractor). These models also track the behavior well for a trajectory that starts off the attractor. The LASSO models both have many terms that are small in magnitude but still nonzero. The LASSO model with a low sparsity parameter has similar performance to the STLSQ and SR3 models, but is not a truly sparse model. As the regularization penalty is increased, rather than removing the unimportant terms in the dynamics the method removes many of the true coefficients in the Lorenz model. The LASSO model with a high sparsity parameter has a very poor fit for the dynamics.

\section{Simultaneous Sparse Inference and Data Trimming}\label{sec:trimming}
Many real world data sets contain corrupted data and/or outliers, which is problematic for model identification methods. 
For SINDy, outliers can be especially problematic, as derivative computations are corrupted.  Many data modeling methods have been adapted to deal with corrupted data, resulting in ``robust'' versions of the methods (such as robust PCA). The SR3 algorithm for SINDy can be adapted to incorporate trimming of outliers, providing a robust optimization algorithm for SINDy.
Starting with least trimmed squares~\cite{rousseeuw1984least}, extended formulations that 
simultaneously fit models and trim outliers are widely used in  statistical learning. Trimming has proven particularly useful in the high-dimensional setting when used with the LASSO approach and its extensions~\cite{yang2015robust,yang2018general}.

The high-dimensional trimming extension applied to~\eqref{eq:sindy} takes the form 
\begin{align}
&\min_{\Xiv, \vv} \sum_{i=1}^m \frac{1}{2} v_i\| (\dot{\Xv} - \Thetav(\Xv)\Xiv)_i\|^2 + \lambda R(\Xiv) \label{eq:trim-sindy} \\
& \mbox{s.t.} \quad 0 \leq v_i \leq 1, \quad {\bf 1}^T\vv = h, \nonumber
\end{align}
where $h$ is an estimate of the number of `inliers' out of the potential $m$ rows of the system. The set 
$\Delta_h := \{ \vv:0 \leq v_i \leq 1, \quad {\bf 1}^T\vv = h \}$
is known as the capped simplex. 
Current algorithms for~\eqref{eq:trim-sindy}, such as those of~\cite{yang2018general}, 
rely on LASSO formulations and thus have significant limitations (see previous section). Here, we use the SR3 strategy~\eqref{eq:explicit_form} 
to extend to the trimmed SINDy problem~\eqref{eq:trim-sindy}: 
\begin{align}
\min_{\Xiv,  \Wv, \vv \in \Delta_h} &\sum_{i=1}^m \frac{1}{2} v_i\| (\dot{\Xv} - \Thetav(\Xv)\Xiv)_i\|^2 \nonumber \\
&+ \lambda R(\Wv) + \frac{1}{2\nu}\|\Xiv - \Wv\|^2. \label{eq:trim-sindy-relax}
\end{align}
We then use the alternating Algorithm~\ref{ALG:TRIM} to solve the problem. The step size $\beta$ is completely up to the user, 
as discussed in the convergence theory (see Appendix~\ref{sec:si_convergence}). 
The trimming algorithm requires specifying how many samples should be trimmed, which can be chosen by estimating the level of corruption in the data. 
Estimating derivatives using central differences, for instance, %
%
makes derivative estimates on either side of the original corrupted sample corrupt as well, meaning that three times as many samples as were originally corrupted will be bad. Thus trimming will need to be more than the initial estimate of how many samples were corrupted.
%
 Trimming ultimately can help identify and remove points with bad derivative estimates, leading to a better SINDy model fit.

\begin{algorithm}[t]
   \caption{SR3-Trimming Algorithm}
   \label{ALG:TRIM}
\begin{algorithmic}
  \State{Input $\epsilon$, $\beta$, $\Wv^0$, $\vv^0$}
  \State{Initialize $k=0$,  err$=2\epsilon$.}
   \While{err$>\epsilon$}
   \State{$k \leftarrow k+1$}
   \State{
   \(
   \begin{aligned} \Xiv^{k}  &= \displaystyle\argmin_{\Xiv}  \sum_{i=1}^m \frac{v_i}{2} \| (\dot{\Xv} - \Thetav(\Xv)\Xiv)_i\|^2 \\ &+  \frac{\|\Xiv - \Wv^{k-1}\|^2}{2\nu}\end{aligned}
   \)
   }
   \State{$ \Wv^k  =  \prox_{\lambda\nu R} (\Xiv^k)$}
   \State{$ \vv^k  =  \proj_{\Delta_h} \left(\vv^{k-1} - \beta \gv_{\vv}\right)$}
   \State{$(\gv_{\vv})_i =  \|(\dot{\Xv} - \Thetav(\Xv)\Xiv^k)_i\|^2$}
   \State{err $=\|\Wv^k - \Wv^{k-1}\|/\nu + \|\vv^k - \vv^{k-1}\|/\beta$}
   \EndWhile
\end{algorithmic}
\end{algorithm}

\begin{figure}
\centering
\includegraphics[width=85mm]{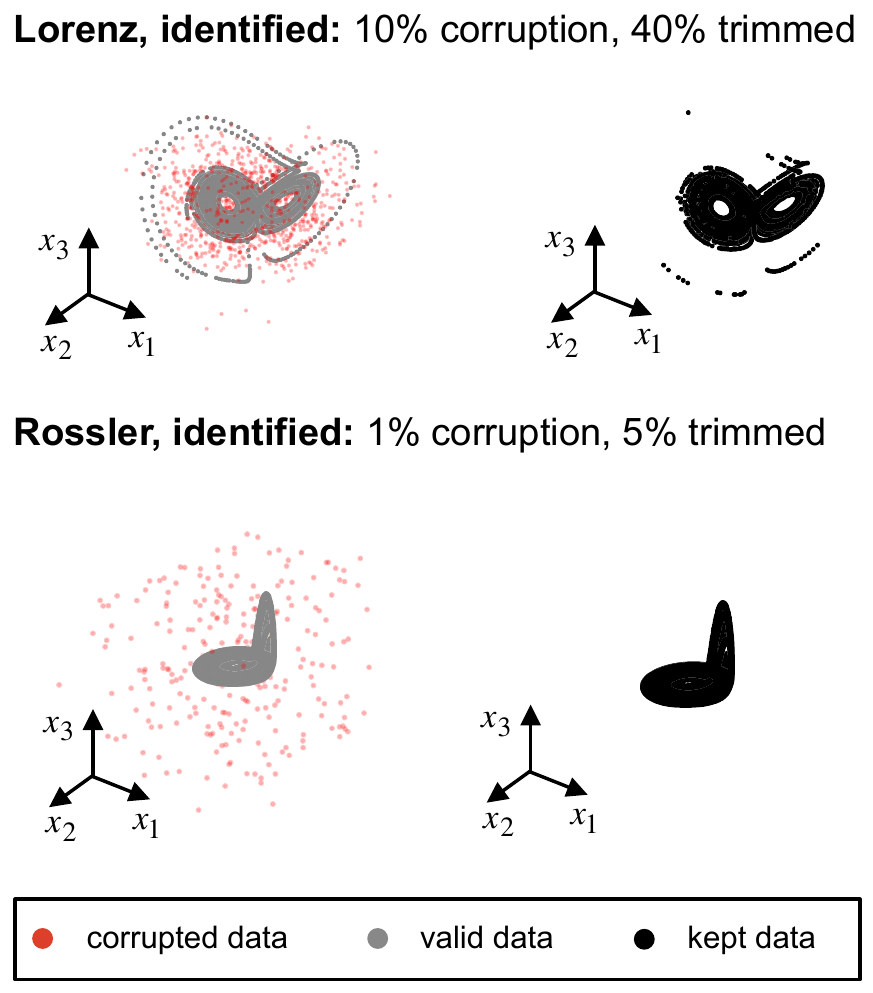}
\caption{Demonstration of the trimming problem for the Lorenz and Rossler systems. For each system, we corrupt some subset of the data (corrupted values shown in red, valid data values shown in gray). We then apply SINDy SR3 with trimming. The black data points show the data that is left after trimming. In both cases, the trimming algorithm correctly identifies the corrupted data points and samples on the attractor remain. For the Lorenz system, some samples from the initial trajectories off of the attractor is also removed.}
\label{fig:trimming}
\end{figure}

\subsection{Example: Lorenz}
We demonstrate the use of SINDy SR3 for trimming outliers on data from the Lorenz system~\eqref{eq:lorenz}. We randomly select a subset of samples to corrupt, adding a high level of noise to these samples to create outliers. We apply the SINDy SR3 algorithm with trimming to simultaneously remove the corrupted samples and fit a SINDy model. Fig.~\ref{fig:trimming} shows the results of trimming on a dataset with 10\% of the samples corrupted. The valid data points are shown in gray and the corrupt data points are highlighted in red. As derivatives are calculated directly from the data using central differences, this results in closer to 30\% corruption (as derivative estimates on either side of each corrupt sample will also be corrupted). We find that the algorithm converges on the correct solution more often when a higher level of trimming is specified: in other words, it is better to remove some clean data along with all of the outliers than to risk leaving some outliers in the data set. Accordingly, we set our algorithm to trim 40\% of the data. Despite the large fraction of corrupted samples, the method is consistently able to identify the Lorenz model (or a model with only 1-2 extra coefficients) from the remaining data in repeated simulations. As initial conditions starting off the attractor are included in the data, some samples of the off-attractor behavior are also trimmed. In contrast, if standard SINDy SR3 (without trimming) is applied to the data set with corrupted samples, the correct model is not identified.

\subsection{Example: Rossler}
The Rossler system
\begin{align}
\dot{x}_1 &= -x_2 - x_3, \nonumber \\
\dot{x}_2 &= x_1 + 0.1 x_2, \label{eq:rossler} \\
\dot{x}_3 &= 0.1 + x_3(x_1 - 14). \nonumber
\end{align}
exhibits chaotic behavior characterized by regular orbits around an attractor in the $x_1,x_2$ plane combined with occasional excursions into the $x_3$ plane. The Rossler attractor is plotted in Fig.~\ref{fig:trimming} with 1\% of samples corrupted (highlighted in red). If standard SINDy SR3 is applied to this data, the system is not correctly identified. However with the algorithm set to trim 5\% of the data, the system is consistently identified (or nearly identified, with only 1-2 incorrect coefficients) in repeated trials.

Much of the Rossler attractor lies in the $x_1,x_2$ plane, which means that the excursions into the $x_3$ dimension can be seen as outliers in the dynamics and are more likely to be flagged as potential outliers. However, if the algorithm is run to convergence most of these points are typically recognized as part of the true dynamics and not removed.

\section{Incorporating physical constraints}\label{sec:constraints}
\begin{figure}
\centering
\includegraphics[width=85mm]{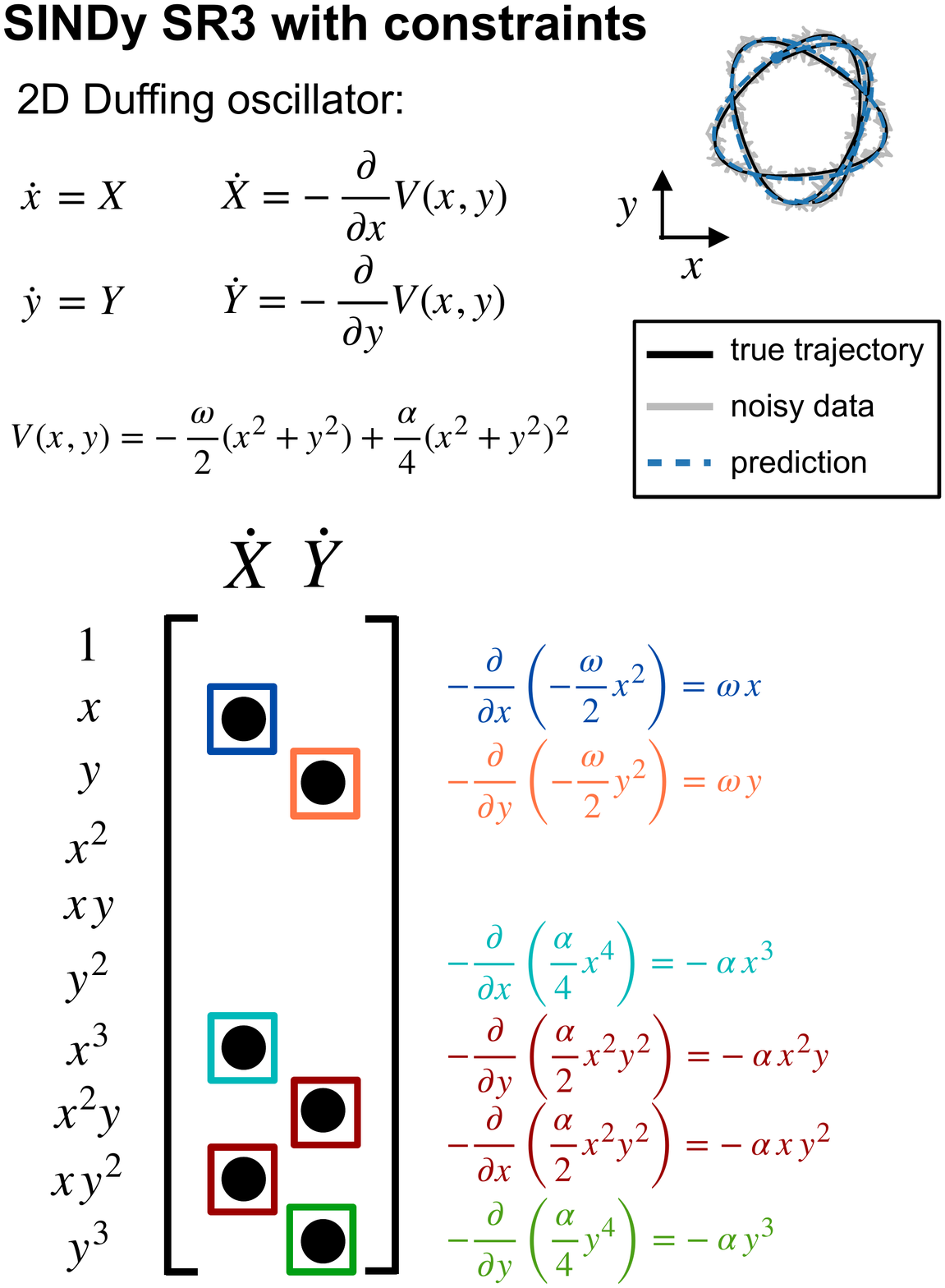}
\caption{Example of using the SINDy SR3 algorithm with constraints. The 2D Duffing oscillator is a Hamiltonian system, where the dynamics are defined by the gradients of a potential function. The gradient requirement imposes a set of constraints on the SINDy model coefficients.}
\label{fig:constraints}
\end{figure}
Physical systems are often subject to constraints based on first principles, such as conservation laws (e.g. conservation of mass or momentum). When performing model discovery, we may require a model that satisfies such a set of desired physical properties. In the case of SINDy, such requirements may manifest as constraints on the form of the coefficient matrix. This constrained approach has already been implemented with the standard thresholded least squares algorithms for SINDy~\cite{Loiseau2017jfm}, and it can also be straightforwardly implemented in the SR3 approach.

\subsection{2D Duffing}
Hamiltonian systems and gradient systems are commonly studied in physics and dynamical systems. In these systems, the equations are gradients of a Hamiltonian or potential function. This imposes a constraint on the form of the coefficient matrix, as the individual equations must be partial derivatives of the same function. We demonstrate the constrained optimization on a Hamiltonian system: the Duffing oscillator in two spatial dimensions. The full system in four variables is
\begin{align}
  \dot{x} &= X \nonumber \\
  \dot{y} &= Y \nonumber \\
  \dot{X} &= - \frac{\partial}{\partial x} V(x,y) \label{eq:2dduffing}\\
  \dot{Y} &= - \frac{\partial}{\partial y} V(x,y) \nonumber
\end{align}
where $x,y$ represent spatial position and $X,Y$ represent momentum. The potential function for the 2D Duffing oscillator is
\begin{equation*}
  V(x,y) = -\frac{\omega}{2}(x^2 + y^2) + \frac{\alpha}{4}(x^2 + y^2)^2.
\end{equation*}
Because the variables $X,Y$ are fixed to represent momentum, we apply SINDy only to fit the second two equations. If $V(x,y)$ is assumed to be a polynomial, the gradient functions will also be polynomials and a specific structure is imposed on the form of the coefficient matrix, with an interdependence between the two columns. In particular, if the equations are written as
\begin{align*}
  \dot{X} &= \xi_{1,1} + \xi_{1,2} x + \xi_{1,3} y + \xi_{1,4} x^2 + \xi_{1,5} x y + \xi_{1,6} y^2 + \cdots \\
  \dot{Y} &= \xi_{2,1} + \xi_{2,2} x + \xi_{2,3} y + \xi_{2,4} x^2 + \xi_{2,5} x y + \xi_{2,6} y^2 + \cdots,
\end{align*}
the constraints that must be imposed are
\begin{align*}
  \xi_{1,3} &= \xi_{2,1} \\
  \xi_{1,5} &= 2 \xi_{2,4} \\
  2 \xi_{1,6} &= \xi_{2,5} \\
  \xi_{1,8} &= 3 \xi_{2,7} \\
  \xi_{1,9} &= \xi_{2,8} \\
  3 \xi_{1,10} &= \xi_{2,9}.
\end{align*}
By vectorizing the coefficient matrix $\xiv = \Xiv(:)$, this set of constraints can be written as a linear constraint $\Cv\xiv = \dv$. The form of the coefficient matrix in the 2D Duffing example is illustrated in Figure~\ref{fig:constraints}.

The problem of interest is now given by 
\begin{equation}
\label{eq:con-sindy} 
\begin{aligned}
&\min_{\Xiv} \frac{1}{2} \| (\dot{\Xv} - \Thetav(\Xv)\Xiv)\|^2 + \lambda R(\Xiv) \\
& \mbox{s.t.} \quad \Cv\xiv = \dv.
\end{aligned}
\end{equation}

Following the relaxation strategy used in the previous sections, we obtain the modified problem 
\begin{equation}
\label{eq:con-sindy-rs}
\begin{aligned}
&\min_{\Xiv, \Wv}  \frac{1}{2} \| (\dot{\Xv} - \Thetav(\Xv)\Xiv)\|^2 + \lambda R(\Wv) + \frac{1}{2\nu} \|\Xiv - \Wv\|^2  \\
& \mbox{s.t.} \quad \Cv\xiv = \dv,  \quad \xiv = \Xiv(:).
\end{aligned}
\end{equation}
To solve this problem, we need to adapt the relax and split strategy to partially minimize the problem in $\Xiv$, which is a quadratic with 
affine equality constraints. This problem has a closed form solution, just as a quadratic problem, through solving an augmented 
system. To form this system, we first use the vec-kron identity:  
\[
\mbox{vec}(\Thetav(\Xv)\Xiv) = (\Iv \otimes  \Thetav(\Xv)) \xiv:=  \Thetav_\Xv^\otimes \xiv
\]
We can form a linear system corresponding to optimality conditions for~\eqref{eq:con-sindy-rs} with respect to $\xiv$: 
\begin{equation}
\label{eq:optCond}
\begin{aligned}
\begin{bmatrix} {\Thetav_\Xv^\otimes}^T\Thetav_\Xv^\otimes + \frac{1}{\nu} \Iv & \Cv^T \\ \Cv & 0 \end{bmatrix}
\begin{bmatrix}\xiv \\ \phiv \end{bmatrix} = 
\begin{bmatrix}
{\Thetav_\Xv^\otimes}^T \dot{\Xv} + \frac{1}{\nu} \Wv^{k-1} \\
\dv
\end{bmatrix}
\end{aligned}
\end{equation}
Conceptually,~\eqref{eq:optCond} is just a large linear system. However, its dimension is very high, and so in the remainder of 
this section we explain how to use the structure of the matrices to solve it efficiently. The most important tools 
include the Kronecker product and associated identities.  We use the fact that
\[
\begin{aligned}
\left({\Thetav_\Xv^\otimes}^T\Thetav_\Xv^\otimes  + \frac{1}{\nu} \Iv\right)^{-1} = \left(\Thetav(\Xv)^T\Thetav(\Xv) + \frac{1}{\nu}\Iv\right)^{-1}\otimes \Iv
\end{aligned}
\]
We can therefore compute the explicit inverse (in the lower dimensional space)
$$\left(\Thetav(\Xv)^T\Thetav(\Xv) + \frac{1}{\nu}\Iv\right)^{-1}$$ and then use it to solve~\eqref{eq:optCond}. 
Specifically, we can reduce~\eqref{eq:optCond} to a small problem 
(same dimension as number of constraints): 
\[
\phiv = \left(\Cv \left(\Thetav(\Xv)^T\Thetav(\Xv) + \frac{1}{\nu}\Iv\right)^{-1}\otimes \Iv \Cv^T \right)^{-1}\mbox{RHS} 
\]  
where the right hand side is given by 
\[
\mbox{RHS}=
\dv - \Cv\left({\Thetav_\Xv^\otimes}^T\Thetav_\Xv^\otimes + \frac{1}{\nu} \Iv\right)^{-1}({\Thetav_\Xv^\otimes}^T \dot{\Xv} + \frac{1}{\nu} \Wv^{k-1}).
\]
Once $\phiv$ is computed, we use~\eqref{eq:optCond} to solve for $\xiv$:
\[
 \xiv = \left({\Thetav_\Xv^\otimes}^T\Thetav_\Xv^\otimes + \frac{1}{\nu} \Iv\right)^{-1}({\Thetav_\Xv^\otimes}^T \dot{\Xv} + \frac{1}{\nu} \Wv^{k-1}  - \Cv^T\phiv)
\]
These steps are summarized in Algorithm~\ref{ALG:CONSTRAINED}.

\begin{algorithm}[t]
   \caption{SR3 Algorithm for Affine Constraints}
   \label{ALG:CONSTRAINED}
\begin{algorithmic}
  \State{Input $\epsilon$, $\Wv^0$}
  \State{Initialize $k=0$,  err$=2\epsilon$.}
   \While{err$>\epsilon$}
   \State{$k \leftarrow k+1$}
   \State{$\Hv_k \leftarrow \left({\Thetav_\Xv^\otimes}^T\Thetav_\Xv^\otimes  + \frac{1}{\nu} \Iv\right)^{-1} $}
   \State{$\rv_k \leftarrow \dv - \Cv\Hv_k({\Thetav_\Xv^\otimes}^T \dot{\Xv} + \frac{1}{\nu} \Wv^{k-1})$}
      \State{$\phiv_k \leftarrow \left(\Cv \Hv_k \Cv^T \right)^{-1}\rv_k $}
      \State{$\xiv_k \leftarrow  \Hv_k({\Thetav_\Xv^\otimes}^T \dot{\Xv} + \frac{1}{\nu} \Wv^{k-1}  - \Cv^T\phiv_k)$}
      \State{$\Xiv_k = \mbox{reshape}(\xiv_k)$}
   \State{$ \Wv^k  =  \prox_{\lambda\nu R} (\Xiv^k)$}
   \State{err $=\|\Wv^k - \Wv^{k-1}\|/\nu$}
   \EndWhile
\end{algorithmic}
\end{algorithm}

We apply both the standard (unconstrained) and constrained SINDy SR3 algorithms to data from the 2D Duffing system with varying levels of noise added. For very low noise levels, both systems correctly identify a coefficient matrix that satisfies the provided constraints. As the noise level increases, the coefficients identified by the unconstrained algorithm drift from satisfying the gradient constraint, while the coefficients identified by the constrained algorithm keep the desired form. In this example it is important to note that the constrained and unconstrained models have comparable performance in reproducing the system behavior via simulation. In other examples, constraints may provide additional benefits such as helping to identify a stable system \cite{Loiseau2017jfm}. Although the constrained SR3 algorithm imposes the constraints on the full coefficients $\Xiv$ and not the auxiliary coefficients $\Wv$, in this example the auxiliary coefficients still satisfy the constraints, producing a model of the desired form.

\section{Parameterized library functions}\label{sec:parameterized}
In standard examples of SINDy, the library is chosen to contain polynomials, which make a natural basis for many models in the physical sciences. However, many systems of interest may include more complicated terms in the dynamics, such as exponentials or trigonometric functions, that include parameters that contribute nonlinearly to the fitting problem. In addition to parameterized basis functions, systems may be subject to parameterized external forcing: for example, periodic forcing where the exact frequency of the forcing is unknown. 
SINDy with unknown parameters is given by 
\vspace{-1mm}
\begin{equation}
\label{eq:sindy-param}
    \min_{\Xiv, \valpha} \frac{1}{2}\| \dot{\Xv} - \Thetav(\Xv, \valpha)\Xiv \|^2 + \lambda R(\Xiv).
\end{equation}
This is a regularized nonlinear least squares problem. The SR3 approach makes it possible to devise an efficient algorithm for this problem as well. 
The relaxed formulation is given by 
\vspace{-1mm}
\begin{equation}
\label{eq:sindy-param-relax}
    \min_{\Xiv, \Wv, \valpha} \frac{1}{2}\| \dot{\Xv} - \Thetav(\Xv, \valpha)\Xiv \|^2 + \lambda R(\Wv)    + \frac{1}{2\nu}\|\Xiv - \Wv\|^2. 
\end{equation}
We solve~\eqref{eq:sindy-param-relax} using Algorithm~\ref{ALG:PARAM}. The $\valpha$ variable is updated using a true Newton step, 
where the gradient and Hessian are computed using algorithmic differentiation. 


\begin{algorithm}[t]
   \caption{SR3-Parameter Estimation}
      \label{ALG:PARAM}
\begin{algorithmic}
  \State{Input $\epsilon$,  $\Wv^0$, $\valpha^0$}
  \State{Initialize $k=0$,  err$=2\epsilon$.}
   \While{err$>\epsilon$}
   \State{$k \leftarrow k+1$}
   \State{
   \(
   \begin{aligned}
     \Xiv^{k}  &= \argmin_{\Xiv}   \frac{1}{2} \| \dot{\Xv} - \Thetav(\Xv, \valpha^{k-1})\Xiv\|^2\\ & +  \frac{1}{2\nu}\|\Xiv - \Wv^{k-1}\|^2
     \end{aligned}
     \)
   }
   \State{$ \Wv^k  =  \prox_{\lambda\nu R} (\Xiv^k)$}
   \State{$\gv^k_{\valpha} = \nabla_{\valpha} \left(\frac{1}{2}\| \dot{\Xv} - \Thetav(\Xv, \valpha)\Xiv^k\|^2\right)$} 
   \State{$\Hv^k_{\valpha} = \nabla^2_{\valpha} \left(\frac{1}{2}\| \dot{\Xv} - \Thetav(\Xv, \valpha)\Xiv^k\|^2\right)$}
   \State{$ \valpha^k  =  \valpha^{k-1} - (\Hv^k_{\valpha})^{-1}\gv^k_{\valpha}$}
   \State{err $=\|\Wv^k - \Wv^{k-1}\|/\nu + \|\valpha^k - \valpha^{k-1}\|$}
   \EndWhile
\end{algorithmic}
\end{algorithm}

\begin{figure}
\centering
\includegraphics[width=\linewidth]{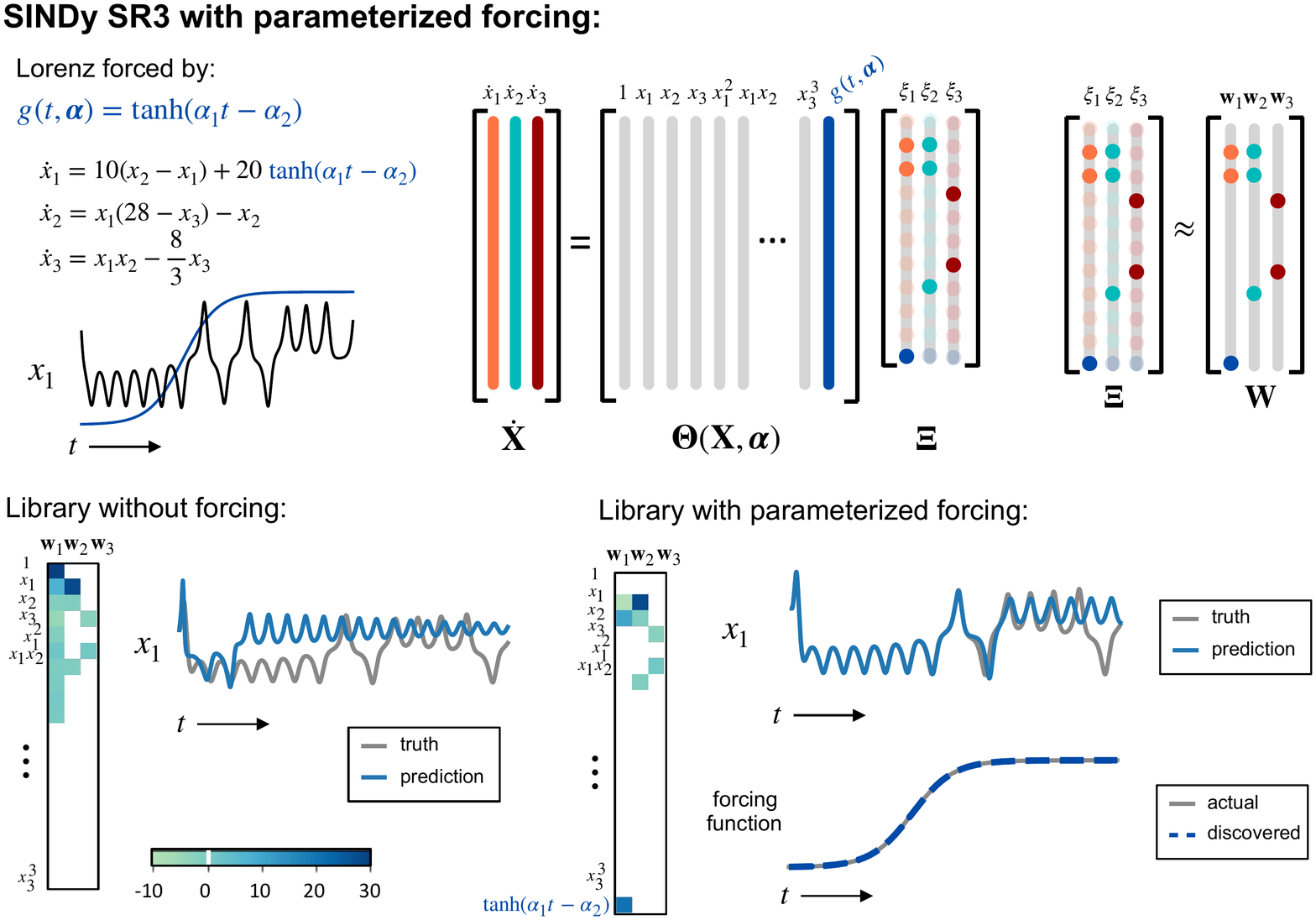}
\caption{Depiction of SINDy SR3 with parameterized library terms, using the example of the Lorenz system forced by a hyperbolic tangent. The library includes a parameterized forcing term and a joint optimization is performed to find the parameter $\alphav$ along with the SINDy model. Without the forcing term, a sparse model is not identified and the resulting model does not reproduce the behavior in simulation. With parameterized forcing in the library, both the forcing parameters and the library can be correctly identified given a sufficiently close initialization of the parameters $\alphav$.}
\label{fig:parameterized}
\end{figure}

The joint optimization for the parameterized case yields a nonconvex problem with potentially many local minima depending on the initial choice of the parameter(s) $\valpha$. 
This makes it essential to assess the fit of the discovered model through model selection criteria.
While the best choice of model may be clear, this means parameterized SINDy works best for models with only a small number of parameters in the library, as scanning through different initializations scales combinatorially with added parameters.

\subsection{Lorenz with parameterized forcing}
We consider~\eqref{eq:lorenz} with $x_1$ forced by a parameterized hyperbolic tangent function $\tanh(\alpha_1 t - \alpha_2)$. The parameters $\alpha_1,\alpha_2$ determine the steepness and location of the sigmoidal curve in the forcing function. We simulate the system  with forcing parameters $\alpha_1=0.8,\alpha_2=3$. Fig.~\ref{fig:parameterized} shows the results of fitting the SINDy model with and without the parameterized forcing term in the library. In the case without forcing, the equation for $x_1$ is loaded up with several active terms in an attempt to properly fit the dynamics. The model is not able to reproduce the correct system behavior through simulation. In the case with forcing, we start with an initial guess of $\alpha_1=5, \alpha_2=10$ and perform the joint optimization to fit both the parameters and the coefficient matrix. The algorithm correctly identifies the forcing and finds the correct coefficient matrix. The resulting system matches the true dynamics for several trips around the attractor.

\section{Discussion}

Machine learning for model discovery in physics, biology and engineering is of growing importance for characterizing complex systems for the purpose of control and technological applications.  Critical for the design and implementation in new and emerging technologies is the ability to interpret and generalize the discovered models, thus requiring that parsimonious models be discovered which are minimally parametrized.   Moreover, model discovery architectures must be able to incorporate the effects of constraints, provide robust models, and/or give accurate nonlinear parameter estimates.  
We here propose the SINDy-SR3 method which integrates a sparse regression framework for parsimonious model discovery with a unified optimization algorithm capable of incorporating many of the critical features necessary for real-life applications.  This includes features for handling corrupt data, imposing constraints, and learning parametrizations, all tasks that the basic SINDy algorithm is unable to handle.
We demonstrate its accuracy and efficiency on a number of example problems, showing that SINDy-SR3 is a viable framework for the engineering sciences.\\

\noindent  All code and data can be found at github.com/kpchamp/SINDySR3

\section*{Acknowledgments}
This material is based upon work supported by the National Science Foundation Graduate Research Fellowship under Grant No. DGE-1256082. SLB acknowledges support from the Army Research Office (W911NF-19-1-0045).  JNK acknowledges support from the Air Force Office of Scientific Research (FA9550-17-1-0329).  Aleksandr Aravkin is supported by the Washington Research Foundation Data Science Professorship.

\newpage

\beginsupplement

\section{Choice of parameters for SR3}\label{sec:si_parameter_choice}
The SR3 algorithm requires the specification of two parameters, $\nu$ and $\lambda$. The parameter $\nu$ controls how closely the relaxed coefficient matrix $\Wv$ matches $\Xiv$: small values of $\nu$ encourage $\Wv$ to be a close match for $\Xiv$, whereas larger values will allow $\Wv$ to be farther from $\Xiv$.

The parameter $\lambda$ determines the strength of the regularization. If the regularization function is the $\ell_0$ norm, the parameter $\lambda$ can be chosen to correspond to the coefficient threshold used in the sequentially thresholded least squares algorithm (which determines the lowest magnitude value in the coefficient matrix). This is because the prox function for the $\ell_0$ norm will threshold out coefficients below a value determined by $\nu$ and $\lambda$. In particular, if the desired coefficient threshold is $\eta$, we can take
\begin{equation*}
\lambda = \frac{\eta^2}{2\nu}
\end{equation*}
and the prox update will threshold out values below $\eta$. In the examples shown here, we determine $\lambda$ in this manner based on the desired values for $\nu,\eta$. If the desired coefficient threshold is known (which is the case for the examples studied here, but may not be the case for unknown systems), this gives us a single parameter to adjust: $\nu$. With $\lambda$ defined in this manner, decreasing $\nu$ provides more weight to the regularization, whereas increasing $\nu$ provides more weight to the least squares model fit.

\section{Simulation details}\label{sec:si_sim}
\subsection{Performance of SR3 for SINDy}\label{sec:si_comparison}
We illustrate a comparison of three algorithms for SINDy using data from the canonical example of the chaotic Lorenz system \eqref{eq:lorenz}. To generate training data, we simulate the system from $t=0$ to 10 with a time step of $\Delta t = 0.005$ for 20 initial conditions sampled from a random uniform distribution in a box around the attractor ($x \in [-36,36],y \in [-48,48], z \in [-16,66]$). This results in a data set with $40 \times 10^4$ samples. We add random Gaussian noise with a standard deviation of $10^{-2}$ and compute the derivatives of the data using the central difference method. The SINDy library matrix $\Thetav(\Xv)$ is constructed using polynomial terms through order 3.

We find the SINDy model coefficient matrix using the following optimization approaches: sequentially thresholded least squares (STLSQ) with threshold $0.1$, SR3 with $\ell_0$ regularization, LASSO with a regularization weight of $0.1$, and LASSO with a regularization weight of $50$. The STLSQ algorithm is performed by doing 10 iterations of the following procedure: (1) perform a least squares fitting on remaining coefficients, (2) remove all coefficients with magnitude less than $0.1$. The LASSO models are fit using the scikit-learn package \cite{scikit-learn}. LASSO models are fit without an intercept. For SR3 we initialize the coefficient matrix using least squares and we use parameters $\nu=1$ and $\lambda=0.005$ (which corresponds to a coefficient threshold of $0.1$, see Appendix~\ref{sec:si_parameter_choice}). For all methods, an unbiasing procedure is performed at the end of the fitting: after the respective optimization algorithm is applied to select the correct terms in the dynamics, a least squares fit is performed on just these terms to find the coefficient values.

For each of the four resulting models we analyze (1) the sparsity pattern of the coefficient matrix and (2) the simulation of the resulting dynamical systems model. We compare the sparsity pattern of the coefficient matrix against the true sparsity pattern for the Lorenz system: SR3 and STLSQ identify the correct sparsity pattern, where as the LASSO models do not. For all models, we simulate the identified system on test trajectories using randomly selected initial conditions from two distributions: near attractor (using same distribution as the training set data) and off attractor (using initial conditions selected in a box outside of the training set data, farther from the attractor). In each case, 50 initial conditions are sampled and the simulations are run from $t=0$ to $5$ for each initial condition. For the near attractor data, the $R^2$ scores for the STLSQ, SR3, and LASSO model with low regularization are $0.82$, $0.82$, and $0.83$, respectively. For the off attractor data, the $R^2$ scores are $0.84$, $0.84$, and $0.82$. In both cases, the LASSO model with heavy regularization had very poor simulation results and had a strongly negative $R^2$ score. For all models, simulations for initial conditions are $(-8,7,27)$ (on attractor) and $(0.01,0.01,60)$ (off attractor) are shown in Figure~\ref{fig:method_comparison} in the main text.

\subsection{Data trimming: Lorenz}\label{sec:si_trimming_lorenz}
We demonstrate the use of the SR3-trimming algorithm \ref{ALG:TRIM} on data from the Lorenz system \eqref{eq:lorenz}. We simulate the system over the same time as in Appendix~\ref{sec:si_comparison} from 5 randomly sampled initial conditions. This results in a data set with $10^4$ samples. We add Gaussian noise to the data with standard deviation $10^{-3}$. We then randomly choose 10\% of the samples to corrupt ($1000$ total samples). For each state variable of each corrupted sample, noise chosen from a random uniform distribution over $[-50,50]$ is added. Derivatives are calculated from the data using central difference after the corruption is applied. We then apply the SR3-trimming algorithm, specifying that around 40\% of the data points will be trimmed. We use SR3 parameters $\nu=1$ and $\lambda=0.005$ (corresponding to a coefficient threshold of $0.1$), and the step size is taken to be the default value $\beta=1$. With repeated testing we find that the algorithm is consistently able to correctly remove the outliers from the data set and identify the Lorenz system.

\subsection{Data trimming: Rossler}\label{sec:si_trimming_rossler}
As an additional example, we test the trimming algorithm on data from the Rossler system \eqref{eq:rossler}. We generate sample data from 5 randomly sampled initial conditions around the portion of the attractor in the $x_1,x_2$ plane, simulating trajectories from $t=0$ to 50 with a time step of $\Delta t = 0.01$. Our data set consists of 25000 samples. We add Gaussian noise with a standard deviation of $10^{-3}$ and add outliers to 1\% of the data in the same manner as in Appendix~\ref{sec:si_trimming_lorenz}, with the noise level chosen from a random uniform distribution over $[-100,100]$. Derivatives are calculated from the corrupted data using central difference. We apply the SR3-trimming algorithm with parameters $\nu=1$ and $\lambda = 1.25 \times 10^{-3}$ (corresponding to a coefficient threshold of $0.05$) and default step size $\beta=1$. On repeated trials we find that if we trim 5\% of the data, the system is correctly identified in most cases (or only 1 or 2 coefficients are misidentified).

\subsection{Physical constraints: 2D Duffing}\label{sec:si_constraints_duffing}
We demonstrate the use of SINDy SR3 with constraints using the example of a 2D Duffing oscillator. We simulate the system~\eqref{eq:2dduffing} from time $t=0$ to 10 with a time step of $0.01$ for 20 initial conditions. The initial conditions are chosen from a uniform distribution with $x,y,X,Y \in [-\pi,\pi]$. We use constrained SINDy SR3 to discover the latter two equations in~\eqref{eq:2dduffing}, imposing the constraints laid out in Section~\ref{sec:constraints}. We use a polynomial library with terms up through order 3 and parameters $\nu=1$ and $\lambda = 0.005$ (corresponding to a coefficient threshold of $0.1$).

To assess the performance of the constrained algorithm, we apply noise at various levels and compare the performance of the constrained algorithm with the unconstrained algorithm. Gaussian noise is applied to the data at standard deviations of 0, 0.01, 0.05, and 0.1. With no noise and noise strength 0.01, both the constrained and unconstrained algorithms discover a model that satisfies the gradient constraint. At stronger noise strengths, both approaches identify the correct terms in the dynamics but only the models found via the constrained approach satisfy the gradient constraint. We also compare performance by simulating the identified systems on test trajectories from 50 randomly chosen initial conditions in the same distribution as the training set initial conditions. While the $R^2$ scores decrease slightly as noise increases, the constrained and unconstrained models have nearly identical scores at all noise levels.

\subsection{Parameterized library functions: Lorenz with parameterized forcing}\label{sec:si_parameterized_lorenz}
To demonstrate the use of SR3 for SINDy with parameter estimation, we look at an example of the Lorenz system \eqref{eq:lorenz} forced by a parameterized hyperbolic tangent function $\tanh(\alpha_1 t - \alpha_2)$. The full set of equations for the system is
\begin{align*}
  \dot{x}_1 &= 10 (x_2 - x_1) + 20 \tanh(\alpha_1 t - \alpha_2) \\
  \dot{x}_2 &= x_1(28 - x_3) - x_2 \\
  \dot{x}_3 &= x_1 x_2 - ({8}/{3}) x_3.  
\end{align*}
The parameters $\alpha_1,\alpha_2$ determine the steepness and location of the sigmoidal curve in the forcing function. We simulate the system as in Appendix~\ref{sec:si_comparison} for a single initial condition $(8,-7,27)$ with forcing parameters $\alpha_1=0.8,\alpha_2=3$. We add Gaussian noise of standard deviation $10^{-3}$ and compute the derivatives via central difference.

We apply Algorithm~\ref{ALG:TRIM} to perform a joint discovery of both the coefficients $\Wv$ and forcing parameters $\alphav$. We use parameters $\nu=0.1$ and $\lambda = 0.05$ (corresponding to coefficient threshold $0.1$). $\Wv$ is initialized using least squares, and as an initial guess for $\alphav$ we use $\alphav^0=(5,10)$. The algorithm discovers the correct parameters $\alphav$ as well as the correct sparsity pattern in the coefficient matrix. We simulate the system and see that the discovered system tracks the behavior for several trips around the attractor. Results are shown in Figure~\ref{fig:parameterized}.

For comparison, we apply the SR3 algorithm for SINDy with no forcing term in the library, using the same SR3 parameters as in the forcing case. The resulting model has many active terms in the equation for $\dot{x}_1$, as it attempts to capture the forcing behavior with polynomials of $x_1,x_2,x_3$. This model does not perform well in simulation, even from the same initial condition used in the training set. Figure~\ref{fig:parameterized} shows the coefficient matrix and model simulation for the discovered system.

\section{Convergence results}\label{sec:si_convergence}
Here we state convergence results for Algorithm~\ref{ALG:BASIC} and Algorithm~\ref{ALG:TRIM}. These algorithms fall under the framework of two classical methods, proximal gradient descent and the proximal alternating linearized minimization algorithm (PALM) \cite{bolte2014proximal}. While we demonstrate the use of Algorithm~\ref{ALG:PARAM} on two example problems, this algorithm is much harder algorithm to analyze due to the complication from the Newton's step. We leave obtaining theoretical guarantees of Algorithm~\ref{ALG:PARAM} as future work.

\subsection{Convergence of Algorithm~\ref{ALG:BASIC}}
Using the variable projection framework \cite{golub1973differentiation}, we partially optimize out $\Xiv$ and then treat Algorithm~\ref{ALG:BASIC} as the classical proximal gradient method on $\Wv$. The convergence result for Algorithm~\ref{ALG:BASIC} is provided in \cite[Theorem 2]{zheng2019unified} and is restated here:
\begin{theorem}
Define the value function as,
\[
p(\Wv) = \min_{\Xiv} \frac{1}{2}\| \dot{\Xv} - \Thetav(\Xv)\Xiv \|^2 + \lambda R(\Wv) + \frac{1}{2\nu}\|\Xiv - \Wv\|^2
\]
When $p$ is bounded below, we know that the iterators from Algorithm~\ref{ALG:BASIC} satisfy,
\[
\frac{1}{N} \sum_{k=1}^{N} \|g^k\|^2 \le \frac{1}{\nu N} (p(\Wv^0) - p^*),
\]
where $g^k \in \partial p(\Wv^k)$ and $p^* = \min_{\Wv} p(\Wv)$.
\end{theorem}

We obtain a sub-linear convergence rate for all prox-bounded regularizers $R$.

\subsection{Convergence of Algorithm~\ref{ALG:TRIM}}
Following the same idea provided by the variable projection framework, the iterations from Algorithm~\ref{ALG:TRIM} are equivalent with an alternating proximal gradient step between $\Wv$ and $\vv$. This is the PALM algorithm, which is thoroughly analyzed in the context of trimming in \cite{aravkin2016smart} and \cite{davis2016asynchronous}. We restate the convergence result here:
\begin{theorem}
Consider the value function,
\begin{align*}
p(\Wv, \vv) = \min_{\Xiv} &\sum_{i=1}^m \frac{1}{2} v_i\| (\dot{\Xv} - \Thetav(\Xv)\Xiv)_i\|^2 \\
& + \lambda R(\Wv) + \frac{1}{2\nu}\|\Xiv - \Wv\|^2
\end{align*}
And we know that the iterators $(W^k, v^k)$ converge to the stationary point of $p$, with the rate,
\[
\min_{k = 0, \ldots, N} \dist(0, \partial p(\Wv^k, \vv^k)) = o\left(\frac{1}{k + 1}\right).
\]
\end{theorem}
Algorithm~\ref{ALG:TRIM} also requires the specification of a step size $\beta$ for the proximal gradient step for $\vv$.
Because the objective is linear with respect to $\vv$, the step size will not influence the convergence result in the above theorem. However, because the objective is non-convex, $\beta$ will have an impact on where the solution lands. In this work we use a default step size of $\beta=1$ for all examples.

\subsection{Convergence of Algorithm~\ref{ALG:CONSTRAINED}}

Algorithm~\ref{ALG:CONSTRAINED} can be analyzed by bootstrapping on the analysis of Algorithm~\ref{ALG:BASIC}. 
As long as the constraint 
\[
\Cv \xiv = \dv
\]
is feasible, it defines an affine subspace $\mathcal{S}$, which we can reparameterize in terms of a new variable
\[
\mathcal{S} = \{\xiv: \Cv \xiv = \dv\} = \xiv_0 + \mbox{Null}(\Cv)
\]
Let $\Sv$ be the basis for the nullspace of $\Cv$. We can now rewrite the value function for~\eqref{eq:con-sindy-rs} in terms of 
 $\Sv$, parametrized by a new variable $\zetav$:
\[
\begin{aligned}
p(\Wv) & = \min_{\zetav} \frac{1}{2}\| \dot{\Xv} - \Thetav(\Xv)(\xiv_0 + \Sv \zetav) \|^2  \\
& + \lambda R(\Wv) + \frac{1}{2\nu}\|(\xiv_0 + \Sv \zetav) - \Wv\|^2
\end{aligned}
\]
Since the solution for $(\phiv, \xiv)$ for the suproblem~\eqref{eq:con-sindy-rs} is unique, 
Algorithm~\ref{ALG:CONSTRAINED} is equivalent to optimizing this value function.

The convergence result for Algorithm~\ref{ALG:BASIC} is therefore also provided by \cite[Theorem 2]{zheng2019unified}, 
where the quadratic term to be used is 
\[
Q(\zetav) = \frac{1}{2}\| \dot{\Xv} - \Thetav(\Xv)(\xiv_0 + \Sv \zetav) \|^2  + \frac{1}{2\nu}\|(\xiv_0 + \Sv \zetav) - \Wv\|^2.
\]

\end{document}